\documentclass[12pt]{article}
\usepackage{amsmath, latexsym, amsfonts, amssymb, amsthm, amscd}
\usepackage{graphics}
\newtheorem{theorem}{Theorem}
\newtheorem{corollary}{Corollary}
\newtheorem{proposition}{Proposition}
\newtheorem{lemma}{Lemma}

\newtheorem{rem}{Remark}

\newcommand{\ind}{{\bf 1}}

\begin{document}
\title{ $F$-divergence minimal equivalent martingale measures and optimal portfolios for exponential Levy models with a change-point}

\maketitle

\begin{center}
{\large S. Cawston}\footnote{$^{,2}$ LAREMA, D\'epartement de
Math\'ematiques, Universit\'e d'Angers, 2, Bd Lavoisier - 49045,
\\\hspace*{.4in}{\sc Angers Cedex 01.}

\hspace*{.05in}$^1$E-mail: suzanne.cawston@univ-angers.fr$\;\;\;$
$^2$E-mail: lioudmila.vostrikova@univ-angers.fr}{\large
 and  L. Vostrikova$^2$}
\end{center}
\vspace{0.2in}

\begin{abstract} 
We study exponential Levy models with change-point which is a random variable, independent from initial Levy
processes. On canonical space with initially enlarged filtration we describe all equivalent martingale measures for
change-point model and we give the conditions for  the existence of f-divergence minimal equivalent martingale measure. Using the
connection between utility maximisation and $f$-divergence minimisation, we obtain a general formula for optimal
strategy in change-point case for initially enlarged filtration and also for progressively
enlarged filtration in the case of exponential utility. We illustrate our results considering the  Black-Scholes model with change-point.
\end{abstract}

%%%%%%%%%%%%%%%%%%%%%%%%%%%%%%%%%%%%%%%%%%%%%%%%%%%%%%%%%%%%%%%%%%%%%%%
\section{Introduction}%%%%%%%%%%%%%%%%%%%%%%%%%%%%%%%%%%%%%%%%%%%%%%%%%
%%%%%%%%%%%%%%%%%%%%%%%%%%%%%%%%%%%%%%%%%%%%%%%%%%%%%%%%%%%%%%%%%%%%%%%

\par The parameters of financial models are generally
highly dependent on time : a number of events (for example the release of information in the press, 
changes in the price of raw materials or the first time a stock price hits some psychological level) can trigger a change in
the behaviour of stock prices. This time-dependency of
the parameters can often be described using a piece-wise constant
function : we will call this case a change-point model. In this context, an
important problem in financial mathematics will be option pricing and hedging. Of
course, the time of change (change-point) for the parameters is not
explicitly known, but it is often possible to make reasonable
assumptions about its nature and use statistical tests for its
detection.
\par Change-point problems have a long history, probably beginning with the
papers of Page \cite{P1}, \cite{P2} in an a-posteriori setting, and of Shiryaev \cite{SH2} in a
 quickest detection setting. The problem was later considered in many papers, see for instance
\cite{D}, \cite{R}, \cite{GR}, \cite{PS}, \cite{BDK}, \cite{SP},
\cite{LO} and also the book \cite{BN} and references there. In
the context of financial mathematics, the question was investigated in
\cite{HZ}, \cite{CHG}, \cite{GUO}, \cite{GJL}, \cite{DE}, \cite{SH4}, \cite{SH5},\cite{KLMM},
\cite{SN} and was often related to a quickest detection approach.
\par It should be noticed that not only quickest detection approach is interesting in financial mathematics, and this fact is related with pricing and hedging of so called default models (see \cite{ABE}, \cite{NMY}  and references there). In mentioned papers a number of very important results was obtained but for the processes without  jump part or with only one jump.
\par The models with jumps, like exponential Levy models, in general, compromise the uniqueness of an
 equivalent martingale measure  when such measure exists. So, one has to choose in some  way an
  equivalent martingale measure to price. Many approaches have been developed and various criteria suggested for this choice of martingale measure, for example risk-minimization in an $L^2$-sense \cite{LT}
\cite{FS}, \cite{S1}, \cite{S2}, Hellinger integrals minimization \cite{CS}, \cite{CSL}, \cite{G}, entropy minimization \cite{MI}, \cite{MIF}, \cite{ES}, $f^q$-martingale measures \cite{JKM} or Esscher measures \cite{HS}.
\par All these approaches can be considered in unified way using so called $f$-divergences, introduced by Ciszar \cite{CZ} and investigated in a number of papers and
books (see for instance \cite{LV} and references there). It should also be noticed that a general characterisation of $f$-divergence minimal martingale measures with applications to exponential Levy models was given first in \cite{GR1}.
\par We recall that for $f$  a  convex function on $\mathbb R^{+,*}$ and two measures $Q$ and $P$ such that $Q<\!< P$, the $f$-divergence of $Q$ with respect to $P$ is defined as 
$$f(Q | P) = \mathbb E _P [f(\frac{dQ}{dP})]$$
where $\frac{dQ}{dP}$ is Radon-Nikodym density of $Q$ with respect to $P$, and $E_P$ is the expectation with respect to
$P$.
We recall that the utility maximisation is closely related to $f$-divergence minimisation via
Fenchel-Legendre transform and this will be one of essential points to obtain an optimal strategy.
\par The aim of this paper is to study  $f$-divergence minimal martingale measures and optimal portfolios from the point of view of utility maximization, for exponential Levy model with
change-point  where the parameters of the model before and after
the change are known and  a change-point  itself is a random variable, independent from initial Levy processes.
We remark that even complete models like Black-Scholes model, become to be incomplete in change-point setting. This is why simple conditioning  with respect to $\tau$ and the use of the results on the processes with independent increments can not give us a right answer immediately.
\par We start by describing our model in more details. We assume the financial market consists of a non-risky asset $B$ with interest rate $(r_t)_{t\geq 0}$, namely
\begin{equation}\label{B}
 B_t= B_0 \exp (\int_0^t r_s\,ds)
\end{equation}
where
\begin{equation}\label{rt}
r_t = r \ind _{\{\tau> t\}}+ \tilde{r} \ind _{\{\tau\leq t\}},
\end{equation}
with $r,\tilde{r}$ interest rates before and after change-point $\tau$, and  a one-dimensional risky asset $S=(S_t)_{t\geq 0}$,
\begin{equation}\label{S}
S_t= S_0 \exp(X_t) 
\end{equation}
where $X$ is a stochastic process obtained by pasting in $\tau$ of two Levy processes $L$ and $\tilde{L}$ together:
\begin{equation}\label{X}
X_t= L_t \ind _{\{\tau> t\}} + (L_{\tau} + \tilde{L}_t - \tilde{L}_{\tau})\ind _{\{\tau\leq t\}}
\end{equation}
Here and further  $L$ and $\tilde{L}$  supposed to be independent Levy processes with characteristics  $(b,c, \nu) $ and  $(\tilde{b},\tilde{c}, \tilde{\nu})$ respectively  which are independent from $\tau$ ( for more details see \cite{Sa}). To avoid unnecessary complications we assume up to now that for change-point model $r$ and $\tilde{r}$ in (\ref{rt}) are equal to zero, and that $S_0=1$.
\par To describe a probability space on which the process $X$ is well-defined, we consider $(D, \mathcal G, \mathbb G)$  the canonical space of right-continuous functions with left-hand limits equipped with its natural filtration $\mathbb G = (\mathcal G_t)_{t\geq 0}$ which satisfies standard conditions: it is right-continuous, $\mathcal G_0= \{\emptyset, D\},\, \bigvee_{t\geq 0}\mathcal G_t = \mathcal G$ . On the product of such canonical spaces we define  two independent Levy processes $L=(L_t)_{t\geq 0}$  and  $\tilde{L}=(\tilde{L}_t)_{t\geq 0}$ with characteristics  $(b,c, \nu) $ and  $(\tilde{b},\tilde{c}, \tilde{\nu})$ respectively and denote by $P$ and $\tilde{P}$ their respective laws which are assumed to be locally equivalent: $P\stackrel{loc}{ \sim }\tilde{P}$.
As we will consider the market on a fixed finite time interval, we are really only interested in $P|_{\mathcal{G}_T}$ and $\tilde{P}|_{\mathcal{G}_T}$ for a fixed $T\geq 0$ and the distinction between equivalence and local equivalence
does not need to be made.
\par Our change-point will be represented by an independent random variable $\tau$ of law $\alpha$ taking values in $([0,T], \mathcal B([0,T])$. The set $\{\tau=T\}$ corresponds to the situation when the change-point does not take place, or at least not on the interval we are studying.
\par On the probability space $(D\times D\times [0,T],\, \mathcal G \times \mathcal G \times  \mathcal B([0,T],\, P\times \tilde{P}\times \alpha )$ we define a measurable map $X$ by (\ref{X}) and we denote by $\mathbb P$ its law. In what follows we use $\mathbb E$ mainly for the expectation with respect to $\mathbb P$ but this notation will be also used for the expectation with respect to $P\times \tilde{P}\times \alpha $.
\par From point of view of observable processes we can have the following situations. If we observe only the process $X$ then the natural probability space to work is $(D, \mathcal G , \mathbb P)$ equipped with the right-continuous version of the natural filtration $\mathbb G = (\mathcal G _t)_{t\geq 0}$ where $ \mathcal G_t = \sigma \{ X_s,\,\, s\leq t\}$ for $t\geq 0$.
Now, if we observe not only the process $X$ but also some complementary variables related with $\tau$ then we can take it in account by the enlargement of the filtration.
First  we consider the filtration $\mathbb{H}$ given by $\mathcal{H}_t=\sigma(\ind_{\{\tau \leq s\}},s\leq t)$ and note that $\mathcal{H}_T=\sigma(\tau)$. Then
we  introduce two filtrations: the initially enlarged filtration $\mathbb F=(\mathcal F_t)_{t\geq 0}$ 
 \begin{equation}
 \mathcal F _0 = \mathcal G_0\vee \mathcal H_T,\,\,\, \mathcal F_t= \bigcap _{s>t}(\mathcal G _s\vee \mathcal H_T)
 \end{equation}
  and the progressively enlarged filtration
 $\hat{\mathbb F}=  (\hat{\mathcal F _t})_{t\geq 0}$ which satisfies :
 \begin{equation}
 \hat{\mathcal F _0 }= \mathcal G_0\vee \mathcal H_0,\,\,\, \hat{\mathcal F_t} = \bigcap _{s>t}(\mathcal G _s\vee \mathcal H_s)
 \end{equation}
In the case of additional information the most natural filtration from the point of view of observable events would be $\hat{\mathbb{F}}$. 
However, it is not so easy to obtain the explicit formulas of optimal strategies for progressively enlarged filtration. So, we start by investigation of optimal strategies for initially enlarged filtration. In a special case of exponential utility it gives us  an optimal strategy for progressively enlarged filtration. Our approach applied for density process with respect to progressively enlarged filtration gives also answer on optimal strategy for logarithmic utility, but a general case is still form an open question.

\par The paper is organized in the following way. In 2. we start by recalling in unified way the facts about f-divergence minimal equivalent martingale measures for exponential Levy models. This information will be used for investigation of change-point case.
\par In 3 we investigate the change-point case. On mentioned probability space and for initially enlarged filtration we describe first all equivalent martingale measures. Then, we introduce as hypotheses, such properties of $f$-divergence minimal equivalent  martingale measures as a preservation of Levy property and a scaling property. The question of preservation of Levy property was considered in details in \cite{CV} and it was shown that the class of $f$-divergences preserving Levy property is larger then  common $f$-divergences, i.e. the functions such that $f''(x)=ax^{\gamma}$, $a>0, \gamma\in\mathbb R$. We recall that 
 these functions are those for which there exists $A>0$ and real $B$, $C$ such that 
$f(x)=Af_{\gamma}(x)+Bx+ C$
where
\begin{equation}\label{formoff}
f_{\gamma}(x)=\begin{cases}& c_{\gamma} x^{\gamma+2} \text{ if }\gamma\neq -1,-2,
\\& x\ln(x) \text{ if }\gamma=-1,
\\&-\ln(x) \text{ if }\gamma=-2.
\end{cases}
\end{equation}
and $c_{\gamma}= \mbox{sign}[(\gamma +1)\,(\gamma+2)]$. 
The conditions for existence and the expression of Radon-Nikodym density $Z^*_T(\tau )$ of $f$-divergence minimal martingale measure for change-point model is given in Theorem \ref{thc}. Then, in Corollaries \ref{c1} and \ref{c2} we give the corresponding results for common f-divergences and, finally, we  apply the results to Black-Scholes change-point model.

\par In 4. we present first some  facts about utility maximisation and the formulas for optimal strategies of single exponential Levy model. Then  we give a decomposition formula for $f'(Z^*_T(\tau ))$ for  initially enlarged filtration. These decompositions  allow us  via the result of \cite{GR1} to identify optimal strategy (see Theorem \ref{optst1}).  We illustrate these results by  considering again the Black-Scholes model with a change-point.

%%%%%%%%%%%%%%%%%%%%%%%%%%%%%%%%%%%%%%%%%%%%%%%%%%%%%%%%%%%%%%%%%%%%%%%%
\section{$f$-divergence minimal EMM's for exponential Levy  model}%%%%%%
%%%%%%%%%%%%%%%%%%%%%%%%%%%%%%%%%%%%%%%%%%%%%%%%%%%%%%%%%%%%%%%%%%%%%%%%
We start by recalling in unified way the facts about $f$-divergence minimal martingale measures for exponential Levy models. Namely, we will consider common $f$-divergences and we will discuss the preservation of Levy property by $f$-divergence   minimal locally equivalent martingale measures (EMM's), we will mention the expressions for so called Girsanov parameters when we change the initial measure $P$ into $f$-divergence minimal EMM's and also the expression of Radon-Nikodym density of these measures via Girsanov parameters.
\par  Let now  $L=(L_t)_{t\geq 0}$ be 
Levy process with parameters $(b,c, \nu)$ where $b$ is the drift
parameter, $c$ is the diffusion parameter and $\nu$ is the Levy measure, i.e. the measure on $\mathbb R\setminus\{0\}$ which satisfies
\begin{equation}\label{1}
\int_{\mathbb R} (x^2\wedge1) \nu (dx)<+\infty .
\end{equation}
We recall that the characteristic function of $L_t$ for $t\in \mathbb{R}^+$ and $u\in \mathbb{R}$ is given then by:
$$\phi_t(u) = \mathbb{E} e^{i u L_t} = e^{\psi (u) t}$$
and in turn, the characteristic exponent 
$$\psi(u) = iub -\frac{1}{2} c u^2 + \int _{\mathbb{R}}( \exp (iux) -1 - i u h(x)) \nu(dx),$$
where from now on,  $h$ is the truncation function. We set $S=(S_t)_{t\geq 0}$ with
$$S_t= S_0\,\exp (L_t)$$ for our risky asset and $B=(B_t)_{t\geq 0}$ for non-risky asset with constant interest rate $r$. We will suppose without loss of generality up to now that $S_0=1$ and
$r=0$.
\par Let $T$ be a fixed horizon and $\mathbb G = (\mathcal G_t)_{t\geq 0}$ be natural filtration. We recall that for  a  convex function $f$ on $\mathbb R^{+,*}$, the $f$-divergence of the restriction $Q_T$ of the measure $Q$ with respect to the restriction $P_T$ of the measure $P$ to $\mathcal G_T$ is: 
$$f(Q_T | P_T) =  E _P [f(\frac{dQ_T}{dP_T})]$$
Here  by convention we set this integral  equal to $+\infty$ if the corresponding function is not integrable.
We recall that $Q_T^*$ is an $f$-divergence minimal equivalent martingale measure if $f(Q_T^* | P_T) < +\infty$ and
$$f(Q_T^* | P_T) = \inf _{Q\in \mathcal M (P)} f(Q_T | P_T)$$
where  $\mathcal M(P)$ is the set of locally equivalent martingale measures supposed to be non-empty.
We also recall that an $f$-divergence minimal equivalent martingale measure $Q^*$ is invariant under scaling  if for all $x\in \mathbb R^{+,*}$
$$f(xQ_T^* | P_T) = \inf _{Q\in \mathcal M (P)} f(xQ_T | P_T)$$
It is called time-invariant if $Q^*$ is the same for all $T>0$.
For a given exponential Levy model $S=S_0\,e^L$, we say that an $f$-divergence minimal martingale measure $Q^*$ preserves the Levy property if $L$ remains a Levy process under $Q^*$. 
\par We recall that the density $Z$ of any equivalent to $P$ measure $Q$ can be written in the form $Z=\mathcal{E}(M)$ where $\mathcal{E}$ denotes the Doleans-Dade exponential and $M=(M_t)_{t\geq 0}$ is a local martingale. It follows from Girsanov theorem  that there exist predictable functions $\beta $  and $Y$ verifying the following integrability conditions : for $t\geq 0$ ($P$-a.s.)
$$ \int_0^{t} \beta _s ^2 ds < \infty , $$ 
$$ \int_0^t \int_{\mathbb{R}}|\,h(y)\,(Y_s(y)-1)\,|\nu^{X, P}(ds, dy) < \infty , $$
and such that
\begin{equation}\label{mart}
M_t=  \int_0^t \beta_s dX^{c}_s+\int_0^t \int_{\mathbb{R}}(Y_s(y)-1)(\mu^{X}-\nu^{X, P})(ds, dy)
\end{equation}
where  $\mu^{X}$ is a  jump measure  of the process $X$ and $ \nu^{X, P}$ is its compensator with respect to $(P,\mathbb G)$, $\nu^{X, P}(ds, dy)= ds \,\nu(dy)$ ( for more details see \cite{JSH}). We will refer to $(\beta,Y)$ as the Girsanov parameters of the change of measure from $P$ into $Q$. It is known from Grigelionis result \cite{GREG} that a semi-martingale is a process with independent increments under $Q$ if and only if their semi-martingale characteristics are deterministic, i.e. the Girsanov parameters do not depend on $\omega$, i.e. $\beta$ depends only on time $t$ and $Y$ depends on $(t,x)$ time and jump size. Since Levy process is homogeneous process, it implies that $X$ will remain a Levy process under $Q$ if and only if there exists $\beta\in\mathbb{R}$ and a positive measurable function $Y$ such that for all $t\leq T$ and all $\omega$, $\beta_t(\omega)=\beta$ and $Y_t(\omega,y)=Y(y)$. \\
 We recall that if Levy property is preserved, $S$ will be a martingale under $Q$ if and only if 
\begin{equation}\label{drift}
b+\frac{1}{2}c+c\beta+\int_{\mathbb{R}}[(e^y-1)Y(y)-h(y)]\nu(dy)=0
\end{equation}
This follows again from Girsanov theorem and reflects the fact that under $Q$ the drift of $S$ is equal to zero. 

As it was mentioned, our aim in this section is to consider in more detail the class of minimal martingale measures for the functions which satisfy (\ref{formoff}).
In particular, the minimal measure for $f$ will be the same as that for $f_{\gamma}$. Minimal measures for the different functions $f_{\gamma}$ have been well studied (see \cite{GR1} and further references). It has  been shown in \cite{K1}, \cite{ES}, \cite{JKM} that in all these cases, the minimal measure, when it exists, preserves the Levy property.\\ 
 Sufficient conditions for the existence of a minimal measure and an explicit expression of the associated Girsanov parameters have been given in the case of relative entropy in \cite{MIF},\cite{HS} and for power functions in \cite{JKM}. It was also shown in \cite{HS} that these conditions are in fact necessary in the case of relative entropy or for power functions. Our aim in this section is to give a unified expression of such conditions for all functions which satisfy $f''(x)=ax^{\gamma}$ and to show that, under some conditions, they are necessary and sufficient.
 We have already mentioned that $f$-divergence minimal martingale measures play an important role in the determination of utility maximising strategies. In this context, it is useful to have further invariance properties for the minimal measures such as scaling and time invariance properties. This is  the case when $f''(x)=ax^{\gamma}$. \\

\begin{theorem}\label{exist}
Consider a Levy process $X$ with characteristics $(b,c,\nu)$  and let $f$ be a function such that $f''(x)=ax^{\gamma}$, where $a>0$ and $\gamma\in\mathbb{R}$. Suppose that $c\neq 0$ or $\stackrel{\circ}{supp}(\nu)\neq \emptyset$. Then there exists an $f$-divergence minimal equivalent  to $P$ martingale measure $Q$ preserving Levy properties if and only if there exist  constants $\alpha, \beta\in\mathbb{R}$ and measurable function  $Y : \mathbb R  \setminus \{0\}\rightarrow \mathbb R^ {+}$  such that
\begin{equation}\label{Y1}
Y(y)=(f')^{-1}(f'(1)+ \alpha(e^{y}-1))
\end{equation}
and such that the following properties hold:
\begin{equation}\label{cdsec1}
Y(y)> 0 \,\,\,\nu-a.e.,
\end{equation}
\begin{equation}\label{cdsec2}
 \int_{|y|\geq 1}(e^{y}-1)Y(y)\nu(dy)<+\infty.
\end{equation}
\begin{equation} \label{cdsec3}
b+\frac{1}{2}c+c\beta+\int_{\mathbb{R}}((e^y-1)Y(y)-h(y))\nu(dy)=0.
\end{equation}
If such a measure exists the Girsanov parameters associated with $Q$ are: $(\beta ,Y)$ if $c\neq 0$, and
$(0 ,Y)$ if $c=0$. In addition, this measure is scale and time invariant.
\end{theorem}
We begin with some technical lemmas. For $Q\stackrel{loc}{\sim}P$ we denote by $(Z_t)_{t\geq 0}$ Radon-Nikodym density process of $Q$ with respect to $P$.\\

\begin{lemma}\label{measures}
Let $Q$ be the measure preserving Levy property. Then, $Q_T\sim P_T$ for all $T>0$ iff
\begin{equation}
Y(y)> 0 \,\,\,\nu-a.e.,
\end{equation}
\begin{equation}\label{hellinger}
 \int_{\mathbb{R}}(\sqrt{Y(y)}-1)^2\nu(dy)<+\infty .
\end{equation}
\end{lemma}

\it Proof \rm \, See Theorem 2.1, p. 209 of \cite{JSH}.$\Box$\\

\begin{lemma}\label{integrability}
Under $Q_T\sim P_T$, the condition $E_P | f(Z_T)| < \infty$ is equivalent to
\begin{equation}\label{predictable}
 \int_{\mathbb{R}} [f(Y(y))-f(1)-f'(1)(Y(y)-1)] \nu(dy)<+\infty 
\end{equation}
\end{lemma}

\it Proof\rm\, In our particular case, $E_P | f(Z_T)| < \infty$ is equivalent to the existence of $E_P  f(Z_T)$. 
We use Ito formula to express this integrability condition in predictable terms. Taking for $n\geq 1$ stopping times 
$$ s_n= \inf\{ t\geq 0 : Z_t> n \,\mbox{or}\, Z_t < 1/n\}$$
where $\inf\{\emptyset\}= +\infty$,
we get for $\gamma\neq -1,-2$ and $\alpha = \gamma +2$ that $P$-a.s.
$$Z_{T\wedge s_n}^{\alpha} = 1 + \int _0^{T\wedge s_n}\alpha\,\beta\,Z_{s-}^{\alpha} d X_s^c 
 + \int _0^{T\wedge s_n}\int_{\mathbb R } Z_{s-}^{\alpha}(Y^{\alpha}(y) -1) (\mu^X - \nu^ {X,P})(ds , dy) $$
$$+ \frac{1}{2 }\alpha\,(\alpha -1)\,\beta^2 \,c\int _0^{T\wedge s_n}\,Z_{s-}^{\alpha} ds  + \int _0^{T\wedge s_n}\int_{\mathbb R } Z_{s-}^{\alpha}[Y^{\alpha}(y) -1 - \alpha(Y(y) -1)] d s\,\, \nu (dy)$$
Hence,
\begin{equation}\label{formula}
Z_{T\wedge s_n}^{\alpha} = \mathcal E( N^{(\alpha )} + A^{(\alpha )})_{T\wedge s_n}
\end{equation}
where
$$N^{(\alpha )}_t=  \int _0^t\alpha\,\beta d X_s^c +
\int _0^t(Y^{\alpha}(y) -1)  (\mu^X - \nu^ {X,P})(ds , dy)$$
and
$$A^{(\alpha )}_t= \frac{t}{2 }\alpha\,(\alpha -1)\,\beta^2 c + t\int_{\mathbb R }[Y^{\alpha}(y) -1 - \alpha(Y(y) -1)]   \nu (dy)$$
Since $[N^{(\alpha )}, A^{(\alpha )}]_t =0$ for each $t\geq 0$ we have
$$Z_{T\wedge s_n}^{\alpha} = \mathcal E( N^{(\alpha )})_{T\wedge s_n}\mathcal E ( A^{(\alpha )})_{T\wedge s_n}$$
In the case $\alpha >1$ and $\alpha <0,$ and  $E_P Z_T^{\alpha}< \infty$,  we have by Jensen inequality
$$0 \leq Z_{T\wedge s_n}^{\alpha} \leq E_P (Z_{T}^{\alpha}\,|\,\mathcal F _{T\wedge s_n})$$
and since the right-hand side of this inequality form uniformly integrable sequence, $(Z_{T\wedge s_n}^{\alpha})_{n\geq 1}$
is also uniformly integrable.
We remark that  $A^{(\alpha )}_t\geq 0$ for all $t\geq 0$ and
$$\mathcal E ( A^{(\alpha )})_{T\wedge s_n} = \exp ( A^{(\alpha )}_{T\wedge s_n})\geq 1.$$
It means that $( \mathcal E( N^{(\alpha )})_{T\wedge s_n})_{n\in\mathbb N^*}$ is uniformly integrable
and 
\begin{equation}\label{du1}
E_P(Z_T^{\alpha}) = \exp (A^{(\alpha )}_T)
\end{equation}
If (\ref{predictable}) holds, then by Fatou lemma and since $\mathcal E( N^{(\alpha )})$ is a local martingale
we get
$$E_P(Z_T^{\alpha}) \leq \underline{\lim}_{n\rightarrow \infty}E_P (Z_{T\wedge s_n})\leq \exp (A^{(\alpha )}_T)<\infty$$
For $0<\alpha <1$, we have again
$$Z_{T\wedge s_n}^{\alpha} = \mathcal E(N^{(\alpha )})_{T\wedge s_n}\mathcal E (A^{(\alpha )})_{T\wedge s_n}$$
with uniformly integrable sequence $(Z_{T\wedge s_n}^{\alpha})_{n\geq 1}$.
Since
$$\mathcal E ( A^{(\alpha )})_{T\wedge s_n} = \exp ( A^{(\alpha )}_{T\wedge s_n})\geq \exp ( A^{(\alpha )}_{T}),$$
the sequence
 $( \mathcal E(N^{(\alpha )})_{T\wedge s_n})_{n\in\mathbb N^*}$ is uniformly integrable and
\begin{equation}
E_P(Z_T^{\alpha}) = \exp (A^{(\alpha )}_T).
\end{equation} 
For $\gamma = -2$ we have that $f(x) = x\ln (x)$ up to linear term and 
$$Z_{T\wedge s_n}\ln (Z_{T\wedge s_n}) = $$
$$\int _0^{T\wedge s_n} (\ln (Z_{s-}) +1)Z_{s-}\beta d X_s^c
 +
\int _0^{T\wedge s_n} \int _{\mathbb R }Z_{s-}[\ln (Z_{s-})(Y(y)-1) +Y(y)\,\ln (Y(y)] (\mu^X - \nu^ {X,P})(ds , dy)$$
$$+ \frac{1}{2} \beta ^2 c\,\int _0^{T\wedge s_n} Z_{s-}ds + \int _0^{T\wedge s_n}\int _{\mathbb R } Z_{s-}(\,Y(y)\ln (Y(y)) -Y(y) +1\,) ds \,\nu (dy)$$
Taking mathematical expectation we obtain:
\begin{equation}\label{log}
E_P [Z_{T\wedge s_n}\ln (Z_{T\wedge s_n})] = E_P \int _0^{T\wedge s_n}Z_{s-}[\frac{1}{2}\beta ^2\, c+\int _{\mathbb R } (\,Y(y)\ln (Y(y)) -Y(y) +1\,) \, \nu (dy)]ds
\end{equation}
If $E_P [Z_{T}\ln (Z_{T})]<\infty$, then the sequence $( Z_{T\wedge s_n}\ln (Z_{T\wedge s_n}))_{n\in\mathbb N^*}$ is uniformly  integrable. In addition,
 $E_P(Z_{s-})=1$ and we obtain applying Lebesgue convergence theorem that
\begin{equation}\label{du2}
E_P [Z_{T}\ln (Z_{T})] = T[\frac{1}{2}\beta ^2\, c+ \int _{\mathbb R } (\,Y(y)\ln (Y(y)) -Y(y) +1\,) \nu (dy)]
\end{equation}
and this implies (\ref{predictable}). If (\ref{predictable}), then by Fatou lemma from (\ref{log}) we deduce that
$E_P [Z_{T}\ln (Z_{T})]<~\infty$.\\
For $\gamma =-1$, we have $f(x)= -\ln (x)$ and exchanging $P$ and $Q$ we get:
$$E_P[-\ln (Z_T)]= E_Q [\tilde{Z}_{T}\ln (\tilde{Z}_{T})] = T[\frac{1}{2}\beta ^2\, c +\int _{\mathbb R } (\,\tilde{Y}(y)\ln (\tilde{Y}(y)) -\tilde{Y}(y) +1\,) \nu^Q (dy)]$$ 
where  $\tilde{Z}_T= 1/Z_T$ and $\tilde{Y}(y)= 1/Y(y)$.
 But $\nu ^Q(dy) = Y(y) \nu(dy)$ and, finally,
\begin{equation}\label{du3}
E_P [-\ln (Z_{T})] = \frac{T}{2}\beta ^2\,c + T\int _{\mathbb R } (-\ln (Y(y)) +Y(y) -1) \nu (dy)
\end{equation}  
which implies (\ref{predictable}). Again by Fatou lemma we get from (\ref{predictable}) that $E_P [-\ln (Z_{T})]<~\infty$ $\Box$\\

\begin{lemma}\label{equivalence} If the second Girsanov parameter $Y$ has a particular form (\ref{Y1})
then the condition
\begin{equation}\label{cdsec2b}
 \int_{|y|\geq 1}(e^{y}-1)Y(y)\nu(dy)<+\infty
\end{equation}
implies the conditions (\ref{hellinger}) and (\ref{predictable}).
\end{lemma}
\it Proof\,\rm \,\, We can cut each  integral in (\ref{hellinger}) and (\ref{predictable}) on two parts and integrate on the sets $\{|y|\leq 1\} $ and $\{|y|>1\}$. Then we can use a particular form of $Y$ and conclude easily writing Taylor expansion of order 2. $\Box$\\

\noindent \it Proof of Theorem \ref{exist}\, Necessity \rm  We suppose that there exist $f$-divergence minimal equivalent martingale measure $Q$ preserving Levy property of $X$. Then, since  $Q_T\sim P_T$, the conditions (\ref{cdsec1}),
(\ref{hellinger}) follow from Theorem 2.1, p. 209 of \cite{JSH}. From Theorem 3 of \cite{CV} we deduce that (\ref{Y1}) holds. Then, the condition (\ref{cdsec2}) follows from the fact that $S$ is a martingale under $Q$. Finally, the condition (\ref{cdsec3}) follows from Girsanov theorem since $Q$ is a martingale measure and, hence,  the drift of $S$ under $Q$ is zero.\\ 

\it Sufficiency \rm We take $\beta$ and  $Y$ verifying the conditions (\ref{cdsec1}),(\ref{cdsec2}),(\ref{cdsec3}) and we construct
\begin{equation}\label{mart1}
M_t=  \int_0^t \beta dX^{c}_s+\int_0^t \int_{\mathbb{R}}(Y(y)-1)(\mu^{X}-\nu^{X, P})(ds, dy)
\end{equation}
As known from Theorem 1.33, p.72-73,  of \cite{JSH}, the last stochastic integral is well defined if
$$C(W)= T  \int_{\mathbb{R}}(Y(y)-1)^2I_{\{|Y(y)-1|\leq 1\}} \nu (dy) < \infty ,$$
$$C(W')= T  \int_{\mathbb{R}}|Y(y)-1|I_{\{|Y(y)-1|> 1\}} \nu (dy) < \infty .$$
But the condition (\ref{cdsec2}), the relation (\ref{Y1}) and  Lemma \ref{equivalence} implies (\ref{hellinger}). So, $(Y-1)\in G_{loc}(\mu ^ X)$ and
$M$ is local martingale. Then we take
$$Z_T= \mathcal E( M)_T$$
and this defines the measure $Q_T$ by its Radon-Nikodym density.  Now, the conditions (\ref{cdsec1}),(\ref{cdsec2}) together with the relation (\ref{Y1}) and  Lemma \ref{equivalence} imply (\ref{hellinger}), and, hence,
from Lemma \ref{measures} we deduce that $P_T\sim Q_T$.\\
Since $P_T\sim Q_T$, the Lemma \ref{integrability} gives us the needed integrability condition: $E_P|f(Z_T)|<~\infty $.  
Now, since (\ref{cdsec3}) holds, $Q$ is martingale measure, and it remains to show that $Q$ is indeed
$f$-divergence minimal. For that we take any equivalent martingale measure $\bar{Q}$ and we show that
\begin{equation}\label{ineq}
E_Q f'(Z_T) \leq E_{\bar{Q}} f'(Z_T).
\end{equation}
If the mentioned  inequality holds, the Theorem 2.2 of \cite{GR1}
implies that $Q$ is an $f$-divergence minimal.\\
In the case $\gamma \neq -1,-2$ we obtain from (\ref{formula}) replacing $\alpha$ by $\gamma +1$:
$$Z_T^{\gamma +1} = \mathcal E(N^{(\gamma +1)})_T \,\,\exp(A_T^{(\gamma +1)})$$
and using a particular form of $f'$ and $Y$  we get that for $0\leq t\leq T$
$$N_t^{(\gamma +1)} = (\gamma+1)\,\beta\, \hat{X}_t$$ 
where $\hat{X}$ is a stochastic logarithm of $S$. So, $\mathcal  E(N^{(\gamma +1)})$ is a local martingale and we get
$$E_{\bar{Q}} Z_T^{\gamma +1} \leq \exp(A_T^{(\gamma +1)}) = E_Q Z_T^{\gamma +1}$$
and, hence, (\ref{ineq}).\\
In the case $\gamma = -1$
we prove using again a particular form of $f'$ and $Y$ that
$$f'(Z_T) = E_Q( f'(Z_T)) +  {\beta} \hat{X}_T$$ 
Since $E_{\bar{Q}} \hat{X}_T\leq 0$ we get that
$$E_{\bar{Q}} f'(Z_T) \leq E_Q f'(Z_T)$$
and it proves that $Q$ is $f$-divergence minimal.\\
The case $\gamma =-2$ can be considered in similar way.\\

Finally, note that the conditions which appear in Theorem \ref{exist} do not depend in any way on the time interval which is considered and, hence, the minimal measure always exists and its Girsanov parameters does not depend on $T$. So, the measure $Q^*$ is time invariant. Furthermore,  if $Q^*$ is $f$-divergence minimal, the equality 
$$f(cx)= Af(x) + B x+ C$$
with $A,B,C$ constants, $A>0$,
gives
$$E_P[f(c\frac{d\bar{Q}}{dP})]=A E_P[f(\frac{d\bar{Q}}{dP})]+B +C \geq A E_P[f(\frac{dQ}{dP})]+ B + C=E_P[f(c\frac{dQ}{dP})]$$
and $Q$ is scale invariant. $\Box$
\begin{corollary}\label{c01}The existence of $f$-divergence EMM for power function $f(x)= c_{\gamma}x^{\gamma +2}$,
$\gamma >-1$ or $\gamma < -2$ is equivalent to the existence of a strictly positive minimizer $Y$ for the integral
$$I_1(\beta, Y)= \frac{1}{2}(\gamma +2)(\gamma +1)\beta^ 2 c + \int_{\mathbb R}(Y^{\gamma +2}(y) -1-(\gamma+2)(Y(y)-1))\nu (dy) $$
under the constraint (\ref{cdsec3}) over the set
$$ \mathcal K _1 = \{ (\beta, Y)\,|\,\beta\in \mathbb R, Y \geq 0,\,
\int_{\mathbb R}|\,(e^y-1)Y(y)-h(y)\,|\nu (dy)< \infty, \,I_1(\beta, Y)<\infty \}$$
For $ -1<\gamma<-2$ it is equivalent to the existence of a strictly positive maximizer $Y$ for the same integral over the same $\mathcal K$.
The solution to these optimisation problems can be only of the form:
$$Y^*(y)= \left\{
\begin{array}{lll}
\left(1+(\gamma+1)\beta^* (e^y -1)\right)^{\frac{1}{\gamma +1}}&\mbox{if}&1+(\gamma+1)\beta^*  (e^ y -1)\geq 0,\\
0 & &\mbox{in opposite case,}
\end{array}\right.$$
with $\beta^*$ satisfying (\ref{cdsec3}).
\end{corollary} 
\begin{corollary}\label{c02} For the function $f(x) = x \ln (x)$, the existence of $f$-divergence EMM is equivalent to
the existence of a strictly positive minimizer $Y$ for the integral
$$I_2(\beta, Y) = \frac{1}{2} \beta^2\,c + \int _{\mathbb R} (Y(y) \ln (Y(y)) - Y(y) +1)\nu(dy)$$
under the constraint (\ref{cdsec3}) over the set
$$ \mathcal K _2 = \{ (\beta, Y)\,|\,\beta\in \mathbb R, Y \geq 0,\,
\int_{\mathbb R}|\,(e^y-1)Y(y)-h(y)\,|\nu (dy)< \infty, \,I_2(\beta, Y) < \infty\}$$
The solution to this optimisation problems can be only of the form:
$$Y(y) = \exp (\beta ^*(e^y-1)) $$
with $\beta^*$ satisfying (\ref{cdsec3}).
\end{corollary}
\begin{corollary}\label{c03} For the function $f(x) = -\ln (x)$, the existence of $f$-divergence EMM is equivalent to
the existence of a strictly positive minimizer $Y$ for the integral
$$I_3(\beta, Y) = \frac{1}{2} \beta^2\,c + \int _{\mathbb R} (- \ln (Y(y)) + Y(y) -1)\nu(dy)$$
under the constraint (\ref{cdsec3}) over the set
$$ \mathcal K _3 = \{ (\beta, Y)\,|\,\beta\in \mathbb R, Y \geq 0,\,
\int_{\mathbb R}|\,(e^y-1)Y(x)-h(x)\,|\nu (dx)< \infty, I_3(\beta, Y) < \infty\}$$
The solution to this optimisation problems can be only of the form:
$$Y(y) =(1 - \beta ^*(e^y-1))^{-1} $$
with $\beta ^*$ satisfying (\ref{cdsec3}). 
\end{corollary}
\noindent \it Proof of Corollaries \ref{c01}, \ref{c02}, \ref{c03}\rm \,The proofs of Corollaries can be performed by using Kunh-Tucker theorem to minimize the mentioned integrals under constraints (see \cite{KIT})
%%%%%%%%%%%%%%%%%%%%%%%%%%%%%%%%%%%%%%%%%%%%%%%%%%%%%%%%%%%%%%%%%%%%%%%%
\section{$f$-divergence minimal EMM's for change-point model}%%%%%%%%%%%
%%%%%%%%%%%%%%%%%%%%%%%%%%%%%%%%%%%%%%%%%%%%%%%%%%%%%%%%%%%%%%%%%%%%%%%%

Here we describe all locally equivalent martingale measures (EMMs) for change point model leaving on our probability space equipped with initially enlarged filtration , and  in particular in relation to the sets of EMMs of the two associated Levy models $L$ and $\tilde{L}$. We denote these sets by $\mathcal M(P)$ and $\mathcal M (\tilde{P})$ respectively. 

%%%%%%%%%%%%%%%%%%%%%%%%%%%%%%%%%%%%%%%%%%%%%%%%%%%%%%%%%%%%%%%%%%%%%%%%%
\subsection{EMMs for change-point model}
%%%%%%%%%%%%%%%%%%%%%%%%%%%%%%%%%%%%%%%%%%%%%%%%%%%%%%%%%%%%%%%%%%%%%%%%%
 We assume that the sets  $\mathcal M(P)$ and $\mathcal M (\tilde{P})$ are non-empty. Let $Q\in \mathcal M (P)$ and $\tilde{Q}\in \mathcal M (\tilde{P})$. We introduce the Radon-Nikodym density processes $\zeta=(\zeta_t)_{t\geq 0}$ and $\tilde{\zeta}=(\tilde{\zeta}_t)_{t\geq 0}$ given by
$$\zeta_t= \frac{dQ_t}{dP_t},\hspace{1cm} \tilde{\zeta}_t= \frac{d\tilde{Q}_t}{d\tilde{P}_t}$$
where $Q_t, P_t, \tilde{Q}_t, \tilde{P}_t$ stand for the restrictions of the corresponding measures to the $\sigma$-algebra $\mathcal G_t$.\\
We also introduce for all $t>0$
$$v_t= \displaystyle\frac{d\tilde{P_t}}{d P_t},$$
then
\begin{equation}\label{4a}
 V_t =
 \ind _{[\![0, \tau]\!]}(t) +  \frac{v _t}{ v _{\tau}}\ind _{]\!]\tau , + \infty [\![}(t)
\end{equation}
 We remark that the measure  $\mathbb P$ which is the law of $X$ verify for $t\geq 0$:
$$\frac{d\mathbb P _t}{d\,P_t}= V_t.$$
 
 \par To describe all EMMs leaving on our space  we define  the process $z=(z_t)_{t\geq 0}$ given by  
\begin{equation}\label{zed}
z_t=\zeta_t\ind _{[\![0, \tau]\!]}(t) +  \zeta_{\tau}\frac{\tilde{\zeta _t}}{\tilde{ \zeta _{\tau}}}\ind _{]\!]\tau , + \infty [\![}(t)
\end{equation}
Finally, we consider  the measure $\mathbb Q$  such that
\begin{equation}\label{5a}
 \frac{d\mathbb Q_t}{d\mathbb P_t}= c(\tau)z_t
 \end{equation}
where $c(\cdot)$ is a measurable function $[0,T]\rightarrow \mathbb R^ {+,*}$ with $\mathbb E c(\tau) = 1$.
%%%%%%%%%%%%
\begin{proposition}\label{mm}
A measure $\mathbb{Q}$ is an equivalent martingale measure for the exponential model ( \ref{X}) related to the process $X$
iff its density process has the form (\ref{5a}).
\end{proposition}
%%%%%%%%%%%%
\noindent \it Proof \rm  First we show that the process $Z=(Z_t)_{t\geq 0}$ given by
\begin{equation}\label{Z1}
Z_t= c(\tau)z_t
\end{equation}
is a density process with respect to $\mathbb P$ and that  the process $S=(S_t)_{t\geq 0}$ such that
\begin{equation}
S_t =  e^ {L_t}\ind _{[\![0, \tau]\!]}(t) +  S_{\tau} e^ {\tilde{L}_t - \tilde{L}_{\tau}}\ind _{]\!]\tau , + \infty [\![}(t)
\end{equation}
is a $(\mathbb Q , \mathbb F)$ - martingale.
\par We begin by noticing that if $M, \tilde{M}$ are two strictly positive $(\mathbb P , \mathbb G)$ martingales  on the same filtered probability space and $\tau$ is a stopping time independent of $M$ and $\tilde{M}$, then $N=(N_t)_{t\geq 0}$ such that
$$N_t= c(\tau )\left[ M_t\ind _{[\![0, \tau]\!]}(t) +  M_{\tau}\frac{\tilde{M _t}}{\tilde{M} _{\tau}}\ind _{]\!]\tau , + \infty [\![}(t) \right]$$
is  a $(\mathbb P , \mathbb F)$ martingale. This fact, for example, can be proved by the conditioning with respect to $\mathcal H_T$ and use the facts that  $M$ and $\tilde{M}$ are  $(\mathbb P , \mathbb G)$ martingales.
To show that $Z$ is a $(\mathbb P , \mathbb F)$-martingale, we prove an equivalent fact that $(V_t\,Z_t)_{t\geq 0}$ is a $(P, \mathbb F)$ - martingale. In fact, the relations  (\ref{4a}), (\ref{zed}), (\ref{Z1}) and previous remark  with
$M_t= \zeta_t$ and $\tilde{M}_t= \tilde{\zeta_t}v_t$ gives the result.
Furthermore, taking conditional expectation with respect to $\mathcal H_T$ and using the fact that $\zeta$ and $\tilde{\zeta}$ are density processes independent from $\tau$, we see that $\mathbb E Z_t=1.$
To show that $S=(S_t)_{t\geq 0}$ is $(\mathbb Q , \mathbb F)$-martingale we establish that $(V_t\,Z_t\,S_t)_{t\geq 0}$ is a $(P, \mathbb F)$ - martingale. For this we use the same remark with $M_t=  e^ {L_t}\zeta_t$ and $\tilde{M}_t= v_t\tilde{\zeta}_t  e^ {\tilde{L}_t}$.
\par Conversely, $Z$ is  the density of any equivalent martingale measure if and only if  $(Z_t\,S_t)_{t\geq 0}$ is a $(\mathbb P, \mathbb F)$ - martingale. But the last fact is equivalent to:   for any bounded  stopping time $\sigma$, 
$$\mathbb E( Z_{\sigma}\,S_{\sigma}) = 1.$$
Replacing $\sigma $ by $\sigma \wedge \tau$ in previous expression we deduce that  $(Z_{t\wedge \tau})_{t\geq 0}$ is the density of a martingale measure for  $( e^{L_{t\wedge\tau}})_{t\geq 0}$. In the same way, using the martingale properties of $Z$ we get for any bounded stopping time $\sigma$ that
$$\mathbb E (\frac{Z_{\sigma}\,S_{\sigma}}{Z_{\sigma\wedge \tau}\,S_{\sigma\wedge \tau}})= 1$$
and so $(\frac{Z_{t}}{Z_{t\wedge \tau}})_{t\geq \tau}$ is the density of an equivalent  martingale measure for 
$(e^{\tilde{L}_{t}-\tilde{ L}_{t\wedge \tau}})_{t\geq \tau}$.~$\Box$

%%%%%%%%%%%%%%%%%%%%%%%%%%%%%%%%%%%%%%%%%%%%%%%%%%%%%%%%%%%
\subsection{From  EMM's to $f$-divergence minimal EMM's.}
%%%%%%%%%%%%%%%%%%%%%%%%%%%%%%%%%%%%%%%%%%%%%%%%%%%%%%%%%%%

\par In the following theorem we give an expression for the density of the $f$-divergence minimal EMM's $\mathbb Q^*_T$ with respect to $\mathbb P_T$ in our change-point framework.  We set for $t\in [0,T]$
$$z^*_T(t) = \zeta^*_{t}\,\frac{\tilde{\zeta}^*_T}{\tilde{\zeta}^*_{t}}$$
where $\zeta^*$ and $\tilde{\zeta^*}$ are the densities of the $f$- divergence EMM's $Q^ *$ and $\tilde{Q}^*$ with respect to $P$ and $\tilde{P}$ respectively.\\
We introduce the following hypotheses :\\\\
$(\mathcal H _1):$  The $f$-divergence minimal equivalent martingale measures $Q^ *$ and $\tilde{Q}^*$ relative to $L$ and $\tilde{L}$ exist.\\\\
 $(\mathcal H _2):$  The $f$-divergence minimal equivalent martingale measures $Q^ *$ and $\tilde{Q}^*$  preserve the Levy property and are invariant under scaling. \\\\
 $(\mathcal H _3):$  For all $c>0$ and $t\in [0,T]$, we have 
$$ \sup_{0\leq t\leq T }\mathbb E |z_T^*(t)\,f'(c\,z_T^*(t))\,|<\infty$$
 where $\mathbb E$ is the expectation  with respect to $\mathbb P$.
 %%%%%%%%%%%%%%%%%%%%%%%%%%%%%
 \begin{theorem}\label{thc} Assume that $f$ is a strictly convex function,  $f \in C^1(\mathbb R ^ {+, *})$,  and that $(\mathcal H _1)$, $(\mathcal H _2)$, $(\mathcal H _3)$ hold.
If  the $f$-divergence EMM's $\mathbb Q^*$ for the change-point model (\ref{X}) exists, then
\begin{equation}\label{6}
\frac{d\mathbb Q_T^ *}{d\mathbb{P}_T}=c(\tau) \,z^*_T(\tau)
 \end{equation}
where $c(\cdot)$ is a measurable function $[0,T] \rightarrow \mathbb R^ +$ such that $\mathbb E c(\tau) = 1$.\\
For $c>0$, let
 $$\lambda_t(c) = \mathbb E [\,z_T^*(t)\,f'(c\,z_T^*(t))\,]$$
 and  let $c_t(\lambda)$ be  its right-continuous inverse. 
 \par If in addition there exists $\lambda^*$ such that 
 \begin{equation}\label{12}
 \int_0^T c_{t}(\lambda ^*)d \alpha (t) = 1,
 \end{equation}
 then the $f$- minimal equivalent martingale measure for a change-point model exists and the density $Z_T^*(\tau)$ of $\mathbb Q^*_T$ with respect to $\mathbb P_T$ is equal to
 $$Z^*_T(\tau)= c^*(\tau) \,z^*_T(\tau)$$
 where
 $c^*(t) = c_t (\lambda ^*)$ for $t\in [0,T]$.
\end{theorem}
\begin{corollary}\label{c1}Assume that $f$ is power function, $f(x) = c_{\gamma}x^{\gamma +2}$. Then under $(\mathcal H _1)$ the $f$-divergence EMM  for change-point model exist and $Z^*_T(\tau)= c^*(\tau) \,z^*_T(\tau)$ with
$$c^*(t)=\frac{[\mathbb{E}(\,z^*_T(t)^{\gamma+2}\,)]^{-\frac{1}{\gamma +1}}}{\int_0^T [\mathbb{E}(\,z^*_T(t)^{\gamma+2}\,)]^{-\frac{1}{\gamma +1}}\,d\alpha(t)}$$
\end{corollary}
\begin{corollary}\label{c2}Assume that $f(x) = x\ln(x)$. Then under $(\mathcal H _1)$ the $f$-divergence EMM for change-point model exist and $Z^*_T(\tau)= c^*(\tau) \,z^*_T(\tau)$ with
$$c^*(t)=\frac{e^{-\mathbb{E}(\,z^*_T(t)\ln z^*_T(t)\,)}}{\int_0^T e^{-\mathbb{E}(\,z^*_T(t)\ln z^*_T(t)\,)}\,d\alpha(t)}.$$ 
\end{corollary} 
\begin{corollary}\label{c3}Assume that $f(x) = -\ln(x)$. Then under $(\mathcal H _1)$ the $f$-divergence EMM for change-point model exist and $Z^*_T(\tau)= c^*(\tau) \,z^*_T(\tau)$ with
$c^*(t)=1$.
\end{corollary}

%%%%%%%%%%%%%%%%%%%% 
\begin{rem}  \rm We can also express the factor $c^*(t)$ in terms of 
$f$-divergences of the processes $L$ and $\tilde{L}$.  Namely, one can see easily that
$$\mathbb{E}(\,{z^*_T(t)}^{\gamma+2}\,)= \mathbb{E}(\,{\zeta^*_t}^{\gamma+2}\,)\, \mathbb{E}(\,{\tilde{\zeta ^*}_{T-t}}^{\gamma+2}\,)  $$
and that
$$\mathbb{E}(\,z^*_T(t)\ln z^*_T(t)\,) = \mathbb{E}(\,\zeta^*_t\ln \zeta^*_t\,) + \mathbb{E}(\,\tilde {\zeta}^*_{T-t}\ln \tilde{\zeta}^*_{T-t}\,), $$
$$\mathbb{E}(-\ln z^*_T(t)\,) = \mathbb{E}(\,-\ln \zeta^*_t\,) + \mathbb{E}(\,-\ln \tilde{\zeta}^*_{T-t}\,). $$
In turn, the last quantities can be easily expressed via the corresponding Girsanov parameters using Ito formula as it was done in Lemma \ref{integrability}.
\end{rem}

\noindent \it Proof  of Theorem \ref{thc} \rm
Since $\mathbb Q^*$ is an equivalent martingale measure we have from (\ref{5a}) that
$$f(\mathbb Q^*_T \,|\,\mathbb P_T)= \mathbb E[f(\,c(\tau)\,\zeta_{\tau }\,\frac{\tilde{\zeta}_{T}}{\tilde{\zeta}_{\tau}})]$$
It follows from the independence of $L, \tilde{L}$ and $\tau$ that 
 $$\mathbb E[f(\,c(\tau)\zeta_{\tau}\,\frac{\tilde{\zeta}_{T}}{\tilde{\zeta}_{\tau}})| \tau=t]=
 \mathbb E[f( \,c(t) \,\zeta_{t}\,\frac{\tilde{\zeta}_{T}}{\tilde{\zeta}_{t}})]$$
 Now, the  independence of $L$ and $\tilde{L}$ implies the conditional independence of $\zeta$ and $\tilde{\zeta}$
given $\sigma (L_s, s\leq T)$. Using the  invariance of $f$ under scaling, we see that in order to minimize $f$-divergence, the measure $\mathbb Q$ should be such that $\zeta$ is the density of an $f$- divergence minimal martingale measure for $(e^ {L_t})_{t\geq 0}$ and $\tilde{\zeta}$ the
 density of an $f$- divergence minimal martingale measure for $(e^ {\tilde{L}_t})_{t\geq 0}$.
Hence (\ref{6}) holds.\\
To find $f$- divergence minimal equivalent martingale measure we have to minimize the function
 $$F(c) = \int _0^T\mathbb E[f(\,c(t) z^ *_T(t)\,)] \,d \alpha (t)$$
 over all cadlag functions $c: [0,T] \rightarrow \mathbb R^ {+,*}$ such that $\mathbb E c(\tau) = 1$. For that we consider the linear space $\mathcal L$ of such cadlag functions  $c:[0;T]\rightarrow \mathbb{R}$  with the norm $||c||= \sup _{t\in [0,T]}| c(t) |$ and also the cone of such positive functions. \\ 
We apply Kuhn-Tucker theorem (see \cite{KIT}) to the function
 $$F_{\lambda}(c) = F(c) - \lambda \int_0^T(c(t)-1)d \alpha (t)$$
 with Lagrangian factor $\lambda>0$. We show that  the Frechet derivative $\frac{\partial  F_{\lambda}}{\partial c}$ of  $F_{\lambda}(c) $, defined by
 \begin{equation}\label{9b}
 \lim _{||\delta ||\rightarrow 0} \frac{| F_{\lambda}(c+\delta) -  F_{\lambda}(c)- \frac{\partial  F_{\lambda}}{\partial c}\delta| }{||\delta ||}=0
 \end{equation}
is equal to:
\begin{equation}\label{11a}
\frac{\partial  F_{\lambda}}{\partial c}(\delta) =
 \int _0^T \left(\mathbb E[f'(\,c(t) z^ *_T(t)\,)z^ *_T(t)] - \lambda \right) \delta(t)d \alpha (t)
 \end{equation}
In fact, by the Taylor formula, we have for $\delta\in \mathcal L$ :
 $$F_{\lambda}(c+\delta) -  F_{\lambda}(c)- \frac{\partial  F_{\lambda}}{\partial c}\delta = $$
  $$\int _0^T\mathbb E[(f'(\,(c(t)+\theta(t)\,) z^ *_T(t))-f'(\,c(t) z^ *_T(t)\,))  z^ *_T(t)]\delta (t) d\alpha (t)$$
where $\theta (t)$ is a function which takes values in the interval $[0, \delta (t)]$.
We remark that the modulus of the right-hand side in the previous equality is bounded from above  by:
  $$A_T = \sup _{t\in [0,T]}\mathbb E [\,|\,f'(\,(c(t)+\theta(t)\,) z^ *_T(t))-f'(\,c(t) z^ *_T(t)\,)\, | \,z^ *_T(t)\,]
 \,\, ||\delta ||$$
Since $f'$ is continuous and increasing  and the functions $c$ and $\delta$ are bounded, hypothesis  $(\mathcal H _3)$ implies that $A_T$ is finite. We conclude by Lebesgue's dominated convergence theorem that (\ref{9b}) holds and then (\ref{11a}).\\ 
Then, in order to $\frac{\partial  F_{\lambda}}{\partial c}\delta = 0$ for all $\delta\in \mathcal L$,
it is necessary and sufficient
to take $c$ such that 
$$\mathbb E[\,z^ *_T(t)\,f'(c(t)z^ *_T(t)) ] - \lambda=0\,\,\,\alpha\mbox{-a.s.}$$
Finally, for  each $c>0$ and $t\in [0,T]$ we consider the function
$$\lambda_t(c)= \mathbb E[\,\,z^*_T(t)\,f'(c z^ *_T(t))\,].$$
We see easily that it is increasing in $c$ and that its right-continuous inverse $c_t(\lambda)$ satisfies:
$$\lambda= \mathbb E[\,z^ *_T(t)\,f'(c_t(\lambda) z^ *_T(t))\,]$$
Now, to obtain a minimizer $c^ *$, it remains to find, if it exists, $\lambda^ *$ which satisfies (\ref{12}).$\Box$\\

\noindent \it Proof  of Corollaries \ref{c1}, \ref{c2} and \ref{c3}. \rm
First of all we remark that for common $f$-divergences, the hypothesis $(\mathcal H _1)$ implies $(\mathcal H _2)$ and $(\mathcal H _3)$. Then, we obtain in power case $f(x)= c_{\gamma}x^{\gamma+2}$ that
  $\lambda_t(c)= (\gamma +2)\,c_{\gamma}\, c^{\gamma +1 }\,\mathbb E[ z^ *_T(t)^{\gamma+2}]$. For
$f(x) = x\ln (x)$ we get $\lambda_t(c)=  \mathbb E[ z^ *_T(t)\ln z^ *_T(t)]+\ln c+1$. In the case $f(x)=-\ln x$ we get $\lambda_t(c)= -1/c$. Finally, we write down $c_t(\lambda)$
and we integrate with respect to $\alpha$ to find $\lambda^*$ and the expression of $c^*(t)$.
$\Box$\\

\noindent  \bf{Example}:\,\,\it A change-point Black-Scholes model. \rm
We  apply the previous results  when $L$ and $\tilde{L}$ define Black-Scholes type models. Therefore, we assume that $L$ and $\tilde{L}$ are continuous Levy processes with characteristics $(b,c,0)$ and $(\tilde{b},c,0)$ respectively, $c>0$. As is well known, the initial models will be complete, with a unique equivalent martingale measure which defines a unique price for options. However, in our change-point model the martingale measure is not unique, and we have an infinite set of martingale measures of the form
$$\frac{d\mathbb{Q}_T}{d\mathbb{P}_T}(X) = c(\tau)\,\exp \,(\,\int_0^T \beta_s dX^c_s-\frac{1}{2}\int_0^T \beta^2_s cds\,)$$
where $c(\cdot)$ is a measurable function $[0,T]\rightarrow \mathbb{R}^{+,*}$ such that $\mathbb E[c(\tau)]=1$ and 
$$\beta_s= -\frac{1}{c}\,[\,(b+\frac{c}{2})\,\ind_{[\![0,\tau]\!]}(s)+(\tilde{b}+\frac{c}{2})\,\ind_{]\!]\tau,+\infty[\![}(s)\,]$$ 
 
\par If for example $f(x)=c_{\gamma}x^{\gamma+2}$ with $\gamma\neq -1,-2$ then applying Theorem \ref{thc}, we get 
$$c^*(t)=\frac{e^{-\frac{\gamma+2}{2c}[(b+\frac{c}{2})^2t+(\tilde{b}+\frac{c}{2})^2(T-t)]}}{\int_0^T e^{-\frac{\gamma+2}{2c}[(b+\frac{c}{2})^2t+(\tilde{b}+\frac{c}{2})^2(T-t)]}d\alpha(t)}$$
If $f(x)=x\ln(x)$, then
$$c^*(t)=\frac{e^{-\frac{1}{2c}[(b+\frac{c}{2})^2t+(\tilde{b}+\frac{c}{2})^2(T-t)]}}{\int_0^T e^{-\frac{1}{2c}[(b+\frac{c}{2})^2t+(\tilde{b}+\frac{c}{2})^2(T-t)]}d\alpha(t)}$$
If $f(x)=-\ln(x)$, then $c^*(t)=1$.

%%%%%%%%%%%%%%%%%%%%%%%%%%%%%%%%%%%%%%%%%%%%%%%%%%%%%%%%%%%%%%%%%%%%%%%
\section{Optimal strategies for utility maximization}%%%%%%%%%%%%%%%%%%
%%%%%%%%%%%%%%%%%%%%%%%%%%%%%%%%%%%%%%%%%%%%%%%%%%%%%%%%%%%%%%%%%%%%%%%
We start by recalling some useful basic facts about optimal strategies for utility maximization.
Then some decomposition formulas will be given which permit us to find optimal strategies. We end up by giving the formulas for optimal strategies for utility maximization in change-point setting for both initially and progressively
enlarged filtrations.
\subsection{Some known facts}

In this subsection, we are interested in finding optimal strategies for terminal wealth with respect to some utility functions. More precisely, we assume that our financial market consists of two assets : a non-risky asset $B$, with interest rate $r$, and a risky asset $S$, modelled using the change-point Levy model defined in (\ref{X}). We denote by $\vec{S}=(B,S)$ the price process and by $\vec{\Phi}=(\phi^0,\phi)$ the amount of money invested in each asset. According to usual terminology, a predictable $\vec{S}$-integrable process $\vec{\Phi}$ is  said to be a self-financing admissible strategy if for every $t\in[0,T]$ and $x$ initial capital
\begin{equation}\label{strat1}
\vec{\Phi}_t\cdot \vec{S}_t=x+\int_0^t \vec{\Phi}_u \cdot d\vec{S}_u
\end{equation}
where the stochastic integral in the right-hand side is  bounded from below. Here $\cdot$ denotes the scalar product. We will denote by $\mathcal{A}$ the set of all self-financing admissible strategies.
In order to avoid unnecessary complications, we will assume again that the interest rate $r$ is $0$, so that starting with an initial capital $x$, terminal wealth at time $T$ is 
$$V_T(\phi)=x+\int_0^T \phi_s dS_s$$
Let $u$ denote a strictly increasing, strictly concave, continuously differentiable function on $dom(u)=\{x\in\mathbb{R}|u(x)>-\infty\}$ which satisfies 
$$u'(+\infty)=\lim_{x\to+\infty}u'(x)=0,$$
$$u'(\underline{x})=\lim_{x\to \underline{x}}u'(x)=+\infty$$
where $\underline{x}=\inf\{u\in dom(u)\}$. 
\par We will say that $\phi ^*$ defines an optimal strategy with respect to $u$ if 
$$\mathbb E _P[u(x+\int_0^T\phi^*_s dS_s)]=\sup_{\phi\in\mathcal{A}}\mathbb E _P[u(x+\int_0^T \phi_s dS_s)]$$
As in \cite{K1}, we will say that ${\phi}^*$ is an asymptotically optimal strategy if there exists a sequence of  admissible strategies $({\phi}^{(n)})_{n\geq 1}$ such that
$$\lim_{n\to+\infty}E[u(x+\int_0^{T}{\phi}^{(n)}_sdS_s)]=\sup_{\phi\in\mathcal{A}}E[u(x+\int_0^T \phi_sdS_s)]$$
As known, there is a strong link between this optimization problem and the previous problem of finding $f$- divergence minimal martingale measures. Let $f$ be the convex conjugate function of $u$ :
\begin{equation}\label{ustar}
f(y)=\sup_{x\in \mathbb{R}}\{u(x)-xy\}=u(I(y))-yI(y)
\end{equation}
where $I=(u')^{-1} = -f'$. We recall that in particular 
$$\begin{aligned}
&\text{ if } u(x)=\ln (x) \text{ then } f(x)=-\ln(x)-1,
\\& \text{ if } u(x)=\frac{x^p}{p},p<1 \text{ then } f(x)=-\frac{p-1}{p}x^{\frac{p}{p-1}},
\\& \text{ if } u(x)=1-e^{-x} \text{ then }f(x)=1-x+x\ln(x).
\end{aligned}$$
The following result  gives us the relation between portfolio optimization and f-minimal martingale measures. 
%%%%%%%%%%%%%%%%%%%%%%%%%%
\begin{theorem} \label{tgr}(cf. \cite{GR1})
Let $x\in \mathbb R^ +$ be fixed and $f\in C^1(\mathbb R^{+,*})$.
Let $Q^ *$ be an equivalent martingale measure which satisfies
$$\mathbb E_P| f(\lambda\frac{dQ^ *_T}{dP_T})| < \infty,\,\,\, \mathbb E_{Q^*}| f'(\lambda\frac{dQ^ *_T}{dP_T}) | < \infty$$
for $\lambda$  such that
$$  -\mathbb E_{Q^*} f'(\lambda\frac{dQ^ *_T}{dP_T}) = x.$$
Then, if $Q^ *$ is an f- divergence minimal martingale measure, there exists  a predictable function $\phi^*$  such that $(\int _0^ {\cdot} \phi^*_u d S_u )$ is a $Q^*$-martingale and
\begin{equation}\label{gr}
-f'(\lambda\frac{dQ^ *_T}{dP_T})  = x + \int _0^ T \phi^*_u d S_u 
\end{equation} 
If the last relation holds, then $\stackrel{\rightarrow}{\Phi }= (\phi ^0, \phi)$ with $\phi _t^0 = x + \int _0^ t \phi_u d S_u - \phi _t S_t$ is an   asymptotically
optimal  portfolio strategy. Moreover, if $\underline{x}> -\infty$, this strategy is optimal.
\end{theorem} 

\it Proof \rm The first part of the Theorem is a slight adaptation of \cite {K1}. We do however recall the proof for the reader's ease. We denote $Z_T = \displaystyle\frac{dQ^ *_T}{dP_T}$.\\
 As $f'$ is strictly increasing, continuous and due to imposed integrability conditions, the function $\lambda\mapsto E_{Q^*}[f'(\lambda Z_T)]$ is also strictly increasing and continuous. Furthermore, since $f'= -(u')^{-1}$, we have $\lim_{\lambda\to 0}E_{Q^*}[f'(\lambda Z_T)]=-\infty$ and $\lim_{\lambda\to+\infty}E_{Q^*}[f'(\lambda Z_T)]=-\underline{x}$. Hence, for all $x>\underline{x}$, there exists a unique $\lambda>0$ such that $E_{Q^*}[f'(\lambda Z_T)]=-x$. As $Q^*$ is minimal for the function $x\mapsto f(\lambda x)$, it follows from Theorem 3.1 of \cite{GR1}, that there exists a predictable process ${\phi}^*$ such that
\begin{equation}\label{eqphi1}
-f'(\lambda Z_T)=x+({\phi}^*\cdot S)_T
\end{equation}
and furthermore ${\phi}^*\cdot S$ defines a $Q^*$-martingale. Then, from the definition of the convex conjugate, we have
$$u(x+({\phi}^*\cdot S)_T)=f(\lambda Z_T) -\lambda Z_T\,f'(\lambda Z_T)$$ and, hence,
$$E_P[|u(x+({\phi}^*\cdot S)_T|]\leq E_P|f(\lambda Z_T)| + \lambda E_P[Z_T|f'(\lambda Z_T)|]<\infty.$$
If now $\phi$ denotes any admissible strategy, we have  from $ u(x)\leq f(y)+xy$ for all $x,y\in \mathbb R^{+,*}$ that
$$\begin{aligned}
u(x+(\phi\cdot S)_T)\leq &(x+(\phi\cdot S)_T)\lambda Z_T+f(\lambda Z_T)
\\\leq &(x+(\phi\cdot S)_T)\lambda Z_T+u(x+({\phi}^*\cdot S)_T)+\lambda Z_Tf'(\lambda Z_T).
\end{aligned}$$
Taking expectation, we obtain since $E_P(Z_T\,f'(\lambda Z_T))= -x$, that
$$E_P[u(x+(\phi\cdot S)_T)]\leq E_P[u(x+({\phi}^*\cdot S)_T]+\lambda E_{Q^*}[(\phi\cdot S)_T].$$
Now, under $Q^*$, $(\phi\cdot S)$ is a local martingale, so that $E_{Q^*}[(\phi\cdot S)_T]\leq 0$. Therefore,
$$E_P[u(x+(\phi\cdot S)_T)]\leq E_P[u(x+({\phi}^*\cdot S)_T)].$$
Furthermore, if $\underline{x}>-\infty$, we note that $({\phi}^*\cdot S)_T\geq \underline{x}-x$, so that ${\phi}^*$ defines an admissible strategy, and hence is a $u$-optimal strategy.
\par When  $\underline{x}=-\infty$, we can construct using the definition of $\mathcal A$
a sequence of admissible strategies ${\phi}^{(n)}$ such that for all $0\leq t\leq T$,
$({\phi}^{(n)}\cdot S)_t\geq -n$ and such that
$$\lim_{n\to+\infty}E[u(x+({\phi}^{(n)}\cdot S)_T)]=\sup_{\phi\in\mathcal{A}}E[u(x+(\phi\cdot S)_T)].$$
Finally, ${\phi}^*$ is asymptotically $u$-optimal. $\Box$\\

In the following theorem proved in \cite{CV1} we give a unified  expression of u-optimal strategy for exponential Levy model. We denote by $(\beta ^* , Y^*)$ the Girsanov parameters for changing of the measure $P$ into $Q^*$. We put also
$$\xi_t(x)= E_{Q^*}[f'(xZ_{T-t})Z_{T-t}] $$

\begin{theorem} \label{strat}
Let $u$ be a $\mathcal{C}^3(]\underline{x},+\infty[)$ utility function and $f$ its convex conjugate. Assume there exists an $f$-minimal martingale measure $Q^*$ which preserves the L\'evy property and such that the integrability conditions are satisfied: for all $\lambda >0$ and all compact set $K\subseteq \mathbb R^+$
\begin{equation}\label{integcd}
E_P|f(\lambda Z_T)|<+\infty,\,\,\,\,E_Q|f'(\lambda\,Z_T)|<+\infty,\,\,\,\,\sup_{t\leq T}\sup_{\lambda \in K}E_Q[f''(\lambda Z_t)Z_t]<+\infty.
\end{equation}\\
Then for any fixed initial capital  $x> \underline{x}$, there exists an  asymptotically $u$-optimal  strategy $\phi ^*$. In addition, $\phi^*$ defines a $u$-optimal strategy as soon as $\underline{x}>-\infty$. \\
Furthermore, if $c\neq 0$, we have
$$\phi _s^*=-\frac{\lambda \beta ^* Z_{s-}}{S_{s-}}\,\xi _s(\lambda Z_{s-})$$
where  $\lambda$ is a unique solution to the equation $E_{Q^*}(-f'(\lambda Z_T))=x$.\\
If $c=0$, $\stackrel{\circ}{supp}(\nu)\neq \emptyset$ and it contains zero, and $Y^*$ is not identically 1, then $f''(x)= ax^{\gamma}$ with $a>0$ and $\gamma \in \mathbb R$, and
$$\phi _s^*=-\frac{\lambda \alpha ^* Z_{s-}}{S_{s-}}\,\xi _s(\lambda Z_{s-})$$
where again  $\lambda$ is a unique solution to the equation $E_{Q^*}(-f'(\lambda Z_T))=x$ and  the constant $\alpha ^*$ is related with the second Girsanov parameter $Y^*$ by the formula:
\begin{equation}\label{gamma}
\alpha ^* =\exp (-y_{0})\,Y(y_0)^{\gamma}\, Y'(y_0)
\end{equation}
where $y_0$ is chosen arbitrarily in $\stackrel{\circ}{supp}(\nu)$.\\
\end{theorem}
In the case of classical utilities we obtain the following result.

\begin{proposition}
Consider a L\'evy process $X$ with characteristics $(b,c,\nu)$ and let $f$ be a function such that $f''(x)=ax^{\gamma}$, where $a>0$ and $\gamma\in\mathbb{R}$. Let $u_f$ be its concave conjugate. Assume there exist $\alpha, \beta\in\mathbb{R}$ and a measurable function $Y:\mathbb{R} \setminus \{0\}\rightarrow \mathbb{R}^{+}$ such that
\begin{equation}  \label{Y}
  Y(y)=(f')^{-1}(f'(1)+ \alpha (e^{y}-1))
\end{equation}
 and such that the following properties hold:
\begin{equation}\label{cdsec1}
Y(y)> 0 \,\,\,\nu-a.e.,
\end{equation}
\begin{equation}\label{cdsec2}
 \int_{|y|\geq 1}(e^{y}-1)Y(y)\nu(dy)<\infty.
\end{equation}
\begin{equation} \label{cdsec3}
b+\frac{1}{2}c+c\beta +\int_\mathbb{R}((e^y-1)Y(y)-h(y))\nu(dy)=0.
\end{equation}
Then if $c\neq 0$, there exists an asymptotically optimal strategy $\phi^* $  given by
$$\phi _s ^*=\alpha_{\gamma}(x)\frac{\beta \,\,Z^{\gamma+1}_{s-}}{E_{Q^*}[Z^{\gamma+1}_{s}]\,S_{s-}},$$
where 
\begin{equation}\label{alpha}
\alpha_{\gamma}(x)= a-(\gamma+1)(x+f'(1)).
\end{equation}
If $c= 0$ and $\stackrel{\circ}{supp}(\nu)\neq \emptyset$, then
$$\phi ^* _s=\alpha_{\gamma}(x)\frac{\alpha \,\,Z^{\gamma+1}_{s-}}{E_{Q^*}[Z^{\gamma+1}_{s}]\,S_{s-}},$$
In addition, $\phi ^*$ is optimal as soon as $\gamma\neq -1$.
\end{proposition}

%%%%%%%%%%%%%%%%%%%%%%%%%%%%%%%%%%%%%%%%%%%%%%%%%%%%%%%%%%%%%%%%%%%%%%
\subsection{A decomposition formula for initially enlarged filtration}
%%%%%%%%%%%%%%%%%%%%%%%%%%%%%%%%%%%%%%%%%%%%%%%%%%%%%%%%%%%%%%%%%%%%%%

We use the structure of $\mathbb Q^*$ presented in Theorem \ref{thc} to write down a decomposition formula mentioned in Theorem \ref{tgr} for $f'(\lambda Z_T^*(\tau))$. 
First of all we give the expressions for Girsanov parameters when changing the measure $\mathbb P$
into $\mathbb Q^*$.
\begin{lemma}\label{girs} Let Girsanov parameters of the $f$-divergence minimal equivalent martingale measures $Q^*$ and $\tilde{Q}^*$ are $(\beta ^*,Y^*)$ and $(\tilde{\beta}^*, \tilde{Y}^*)$ respectively.
Then the Girsanov parameters when changing from $\mathbb P$ to $\mathbb Q^*$ are:
$$\beta ^*_t(\tau)= \beta ^* \ind_{[\![0,\tau]\!]}(t)+\tilde{\beta}^*\,\ind_{]\!]\tau,+\infty[\![}(t) $$
$$Y^*_t(\tau)= Y^* \ind_{[\![0,\tau]\!]}(t)+\tilde{Y}^*\,\ind_{]\!]\tau,+\infty[\![}(t). $$ 
\end{lemma}

Next, we introduce for fixed $u\in [0,T]$, $x\geq 0$ and $t\in [0,T]$ the quantities
$$\rho^{(u)} (t,x)= \mathbb E_{\mathbb Q^*} ( f'(Z^*_T(\tau ))\,|\,\tau =u,\, Z^*_t(u)=x)$$
and we remark that
$$\rho^{(u)} (t,x)= \mathbb E_{\mathbb Q_u^*} (f'(Z^*_T(u ))\,|\, Z^*_t(u)=x) $$
where $\mathbb Q_u^*$ is conditional probability $\mathbb Q^*$ given $\tau =u$.
We notice  that  for regular conditional probabilities and for right-continuous versions of conditional expectations we have: $\mathbb P$-a.s. for all $t\in [0,T]$
\begin{equation}
 \mathbb E _{\mathbb Q^*}(\, f'(Z^*_T(\tau ))\,|\, \mathcal F_t) = \rho ^{(\tau)} (t,Z^*_t(\tau ))
\end{equation}
To simplify the notation  we introduce $\eta_{T-t}(u)$ such that
$$\eta_{T-t}(u) = \frac{z^*_T(u)}{z^*_t(u) }$$
As a consequence of previous formulas, we have
$$\rho ^{(u)} (t, x)=\mathbb E [\eta_{T-t}(u) f' (x \eta_{T-t}(u)) ]$$
Now, we would like to use Ito formula  for $\rho^{(u)} (t,Z^*_t(u))$.
But the mentioned function is not sufficiently smooth and we will  proceed  by approximations.
For that we construct a sequence of  functions $(\phi _n)_{n\geq 1}$.
\\
\begin{lemma}\label{approx} Let $f$ be convex function belonging to $C^3(\mathbb R^{+,*})$. There exists a sequence of bounded functions $(\phi_n)_{n\geq 1}$, which are of class $\mathcal{C}^2$ on $\mathbb{R}^{+,*}$, increasing, such that for all $n\geq 1$, $\phi_n$ coincides with $f'$ on the compact set $[\frac{1}{n},n]$ and such that for sufficiently big $n$ the following inequalities hold for all $x,y>0$~:
\begin{equation}\label{apprfprime}
|\phi_n(x)|\leq 4|f'(x)|+ \alpha \text{ , }|\phi'_n(x)|\leq 3f''(x) \text{ , }|\phi_n(x)-\phi_n(y)|\leq 5|f'(x)-f'(y)|
\end{equation}
where $\alpha$ is a real positive constant.
\end{lemma} 

\noindent \it Proof \rm 
We set, for $n\geq 1$, 
$$A_n(x)=f'(\frac{1}{n})-\int_{x\vee \frac{1}{2n} }^{\frac{1}{n}}f''(y)(2ny-1)^2(5-4ny)dy$$
$$B_n(x)=f'(n)+\int_n^{x\wedge (n+1)} f''(y)(n+1-y)^2(1+2y-2n)dy$$
and finally
$$\phi_n(x)=\begin{cases}
& A_n(x) \text{ if }0 \leq x< \frac{1}{n},
\\&f'(x) \text{ if }\frac{1}{n}\leq x\leq n,
\\& B_n(x) \text{ if }x>n.

\end{cases}$$
\it Proof. \rm \, We can verify easily that $\phi _n$ coincide with $f'$ on $[\frac{1}{n}, n]$ and that the properties (\ref{apprfprime}) hold.$\Box$\\
\par Now we replace $f'$ by $\phi_n$ in previous formulas  and we introduce 
$$\rho^{(u)}_n (t,x)= \mathbb E _{\mathbb Q^*} ( \phi_n(Z^*_T(\tau))\,|\,\tau=u,\, Z^*_t(u)=x)$$
It is not difficult to see that
$$\rho^{(u)} _n(t,Z^*_T(u))=\mathbb E [\eta_{T-t}(u) \phi_n (x \eta_{T-t}(u)) ]$$
In the next lemma we give  a decomposition formula for $\rho^{(u)} _n$. For that we put

\begin{equation}\label{xin}
\xi_t^{(n,u)}(x)= \mathbb{E}[\eta ^2_{T-t}(u)\phi'_n(x\eta_{T-t}(u))]
\end{equation}
and
\begin{equation}\label{han}
H_t^{(n,u)}(x,y)=\mathbb{E}(\eta_{T-t}(u)[\phi_n(x\eta_{T-t}(u)Y^*_t(y))-\phi_n(x\eta_{T-t}(u))])
\end{equation}

\begin{lemma}\label{decomplemma1}
We have $\mathbb Q^*_u$-a.s., for all $t\leq T$,
\begin{equation}\label{ndecomp1}
E_{\mathbb Q^*_u}[\phi_n( Z^*_T(u))\,|\,\mathcal G_t]=E_{\mathbb Q^*_u}[\phi_n( Z^*_T(u))] +
\end{equation}
$$\int_0^t \beta^* _s(u) Z^*_{s-}(u ) \xi^{(n,u)}_s ( Z^*_{s-}(u))dX^{(c),\mathbb Q^*_u}_s+\int_0^t \int_{\mathbb{R}}H^{(n,u)}_s( Z^*_{s-}(u),y)\,(\mu^X-\nu ^{X,\mathbb Q^*_u})(ds, dy)$$
where  $\nu ^{X,\mathbb Q^*_u}$ is a compensator of the jump measure $\mu^X$ with respect to $(\mathbb G, \mathbb Q^*_u)$.
\end{lemma}
\noindent \it Proof \rm In order to apply the Ito formula to $\rho^{(u)}_n$, we show that $\rho_n$ is twice continuously differentiable with respect to $x$ and once with respect to $t$ on the set  $x\geq \epsilon$, $\epsilon >0$ and $t\in[0,T]$ and  that the corresponding derivatives are bounded on the mentioned set. 
Then we apply the Ito formula to $\rho^{(u)}_n$ but stopped at stopping times
$$s_m= \inf \{ t\geq 0\,|\, Z^*_t(u) \leq \frac{1}{m} \},$$
with $m\geq 1$ and $\inf\{\emptyset\}= \infty$. 
\par From strong Markov property of Levy processes we have:
$$\rho^{(u)} _n(t\wedge s_m, Z^*_{t\wedge s_m}(u)) = E_{\mathbb Q^*_u}(\phi_n(Z^*_T(u))\,|\,\mathcal G_{t\wedge s_m})$$
and we remark that $( E_{\mathbb Q^*_u}(\phi_n(Z^*_T(u))\,|\,\mathcal G_{t\wedge s_m}))_{t\geq 0}$ is a $\mathbb Q^*_u$- martingale. By Ito formula we obtain that:\\
$\rho^{(u)} _n(t\wedge s_m, Z^*_{t\wedge s_m}(u)) = \rho^{(u)} _n(0, Z^*_{0}(u))+\displaystyle\int _0^{t\wedge s_m}\frac{\partial \rho^{(u)}_n}{\partial s}(s, Z^*_{s-}(u)) ds +$\\
$$\int _0^{t\wedge s_m}\frac{\partial \rho^{(u)}_n}{\partial x}(s, Z^*_{s-}(u)) d Z^*_{s-}(u)+
\frac{1}{2} \int _0^{t\wedge s_m}\frac{\partial ^2 \rho^{(u)}_n}{\partial x^2}(s, Z^*_{s-}(u)) d\langle Z^{*,c} (u)\rangle _s +$$
$\displaystyle\int _0^{t\wedge s_m}\int_{\mathbb R}[ \rho^{(u)} _n(s,Z^*_{s-}(u)+y)-  \rho^{(u)} _n(s,Z^*_{s-}(u)) -\displaystyle\frac{\partial \rho^{(u)}_n}{\partial x}(s, Z^*_{s-}(u))y] \mu^{Z^*}(ds, dy)$\\\\
Then we can write that
$$\rho^{(u)} _n(t\wedge s_m, Z^*_{t\wedge s_m}(u)) = A_{t\wedge s_m } + M_{t\wedge s_m }$$
with for $0\leq t\leq T$
$$A_{t}= \int _0^{t}\frac{\partial \rho^{(u)}_n}{\partial s}(s, Z^*_{s-}(u)) ds +
\frac{1}{2} \int _0^{t}\frac{\partial ^2 \rho^{(u)}_n}{\partial x^2}(s, Z^*_{s-}(u)) d\langle Z^{*,c} (u)\rangle _s +$$
$\displaystyle \int _0^{t}\displaystyle\int_{\mathbb R}[ \rho^{(u)} _n(s,Z^*_{s-}(u)+y)-  \rho^{(u)} _n(s,Z^*_{s-}(u)) -\displaystyle\frac{\partial \rho^{(u)}_n}{\partial x}(s, Z^*_{s-}(u))\,y] \nu^{Z^*, \mathbb Q^*_u}(ds, dy)$\\\\
and
$$M_{\cdot} = \int _0^{t}\frac{\partial \rho^{(u)}_n}{\partial x}(s, Z^*_{s-}(u)) d Z^*_{s-}(u)+$$
$\displaystyle\int _0^{t}\displaystyle\int_{\mathbb R}[ \rho^{(u)} _n(s,Z^*_{s-}(u)+y)-  \rho^{(u)} _n(s,Z^*_{s-}(u))( \mu^{Z^*}(ds, dy)-\nu^{Z^*, \mathbb Q^*_u}(ds, dy))$\\\\
But since $A$ is predictable process and $( E_{\mathbb Q^*_u}(\phi_n(Z^*_T(u))\,|\,\mathcal G_{t\wedge s_m}))_{t\geq 0}$ is a $\mathbb Q^*_u$- martingale, we obtain that $\mathbb Q^*_u$-a.s.,  $A_t=0$ for all $0\leq t \leq T$.
\par From \cite{RY}, corollary 2.4, p. 59, we get since $\sigma (\cup _{m=1}^{\infty} \mathcal G_{t\wedge s_m}) = \mathcal G _t$ that
$$\lim_{m\rightarrow \infty} \rho^{(u)} _n(t\wedge s_m, Z^*_{t\wedge s_m}(u))= E_{\mathbb Q^*}(\phi_n(Z^*_T(u))\,|\,\mathcal G_{t}) $$
Moreover, we remark that for all $x\in \mathbb R$ and $s\in[0,T]$
$$\frac{\partial \rho^{(u)}_n}{\partial x}(s,x) = \xi^{(n,u)}_s(x)$$
and all $x,y\in \mathbb R$ and $s\in[0,T]$
$$H^{(n,u)}_s(x,y)=  \rho^{(u)} _n(s,x\,Y_s^*(y))-  \rho^{(u)} _n(s,x) $$
Using the definition of local martingales we conclude that the decomposition of Lemma holds.
  $\Box $\\
\par The next step consists to pass to the limit in previous decomposition. For that let
 us denote for $0\leq t \leq T$
\begin{equation}\label{xi}
\xi^{(u)}_t(x)= \mathbb{E}[\eta^2_{T-t}(u)f''(x\eta_{T-t}(u))]
\end{equation}
and
\begin{equation}\label{ha}
H^{(u)}_t(x,y)=\mathbb{E}(\eta_{T-t}(u)[f'(x\eta_{T-t}(u)Y^*_t(y))-f'(x\eta_{T-t}(u))])
\end{equation}
\begin{lemma}\label{decomplemma2}
We have $\mathbb Q^*_u$-a.s., for all $t\leq T$,
\begin{equation}\label{ndecomp2}
  E_{\mathbb Q^*_u}(f'(Z^*_T(u))\,|\, \mathcal G_t) = E_{\mathbb Q^*_u}[f'( Z^*_T(u))] +
\end{equation}
$$\int_0^t  \beta ^* _s(u) \,Z^*_{s-}(u )\, \xi^{(u)}_s ( Z_{s-}(u))dX^{(c),\mathbb Q^*_u}_s+\int_0^t \int_{\mathbb{R}}H^{(u)}_s( Z_{s-}(u),y)\,(\mu^X-\nu ^{X,\mathbb Q^*_u})(ds, dy)$$
where  $\nu ^{X,\mathbb Q^*_u}$ is a compensator of the jump measure $\mu^X$ with respect to $(\mathbb F, \mathbb Q^*_u)$.
\end{lemma}
\it Proof. \rm The proof consists to show the convergence in probability of stochastic integrals and conditional expectations using the properties of $\phi _n$ cited in Lemma \ref{approx} and can be performed in the same way as in
\cite{CV1}.$\Box $\\

%%%%%%%%%%%%%%%%%%%%%%%%%%%%%%%%%%%%%%%%%%%%%%%%%%%%%%%%%%%%%%%%%%%%%%%
\subsection{Optimal strategies in a change-point situation for initially enlarged filtration}
%%%%%%%%%%%%%%%%%%%%%%%%%%%%%%%%%%%%%%%%%%%%%%%%%%%%%%%%%%%%%%%%%%%%%%%
Let  $u$ be a utility function belonging to $C^3(]\underline{x}, +\infty[)$ and $f$ its convex conjugate, $f\in C^3(\mathbb R^{+,*})$. We suppose that $\mathcal M(P)\neq \emptyset$ and $\mathcal M(\tilde{P})\neq \emptyset$ and we introduce the following hypotheses
\\
$(\mathcal H _4):$ For each compact set $K$ of $\mathbb R^{+,*}$  we have:
 $$ \sup_{\lambda\in K}\sup_{t\in [0,T]}E _{Q^*} [\zeta_t^* \,f''(\lambda\,\zeta_t^*)\,]<\infty,\,\,\, \sup_{\lambda\in K}\sup_{t\in [0,T]}E_{\tilde{Q^*}}[\tilde{\zeta}^*_t\, f''(\lambda\,\tilde{\zeta}^*_t)\,] < \infty$$\\
where $\zeta^*$ and $\tilde{\zeta^*}$ are the densities of the $f$-divergence EMM's $Q^ *$ and $\tilde{Q}^*$ with respect to $P$ and $\tilde{P}$ respectively.

%%%%%%%%%%%%%%%%%%%%%%%%%
\begin{theorem}\label{optst1} Let $u$ be a  strictly concave function  belonging to $C^3(]\underline{x}, +\infty[)$.
 Suppose that a convex conjugate $f$ of $u$  satisfy
 $(\mathcal H _1)$, $(\mathcal H _2)$, $(\mathcal H _3)$, $(\mathcal H _4)$ and (\ref{12}).
Then for  change-point model (\ref{X}) there exists an $\mathbb{F}$-optimal strategy $\phi ^*$ . If $c\neq 0$, then
\begin{equation}\label{opt}
\phi^ * _t= -\lambda \,\frac{ {\beta _t}^*(\tau)\, Z^* _{t-}(\tau)}{S_{t-}}\,\xi^{(\tau)}_t ( \lambda\,Z^* _{t-}(\tau )) 
\end{equation}
with $\beta^*$ defined in Lemma \ref{girs} and $\lambda$ such that $\mathbb E_{\mathbb Q^*}(-f'(\lambda\,Z_T^*(\tau)))=x$.
\par If $c=0$ and $\stackrel{\circ}{supp}(\nu)\neq \emptyset$, $\stackrel{\circ}{supp}(\tilde{\nu})\neq \emptyset$ both supports containing $0$,and $Y$, $\tilde{Y}$ are not identically 1,  then $f''(x)= ax^{\gamma}$ with $a>0$ and $\gamma\in\mathbb R$, and
the optimal strategies are defined by the same formula but with the replacement of $\beta^*_t(\tau)$ by $\alpha ^*_t(\tau)$ such that 
$$\alpha^*_t(\tau)= e^{-y_0}{Y^*(y_0)}\!^{\gamma}\,\,\frac{dY^*}{d y}(y_0) \ind _{\{\tau> t\}} +  e^{-y_1}{\tilde{Y}^*(y_1)}\!^{\gamma}\, \,\frac{d \tilde{Y}^*}{d y}(y_1)\ind _{\{\tau\leq t\}}$$
with any $y_0\in\stackrel{\circ}{supp}(\nu)$ and $y_1\in \stackrel{\circ}{supp}(\tilde{\nu})$.
\end{theorem}
%%%%%%%%%%%%%%%%%%%%%%%%%
\noindent\it Proof \rm From Theorem \ref{thc} and the hypotheses $(\mathcal H _1)$, $(\mathcal H _2)$, $(\mathcal H _3)$ and (\ref{12}) it follows that there exists an $f$- divergence minimal martingale measure $\mathbb{Q}^*$.  Since the processes $X$ and $S$ are $\mathbb F$-adapted,  applying  Theorem 3.1 in \cite{GR1}, 
we have the existence of an $\mathbb F$-adapted optimal strategy $\phi ^*$ such that
$$-f'(\lambda Z^*_T(\tau))=x+\int_0^T \phi^*_udS_u$$
and such that $\int_0^. \phi ^*_u dS_u$ defines a local martingale with respect to
$(\mathbb Q,\mathbb F)$. 
Then, for $c\neq 0$, we compare the decomposition of Lemma \ref{decomplemma2} and the decomposition of Theorem  \ref{tgr} to get our formulas. For $c=0$  we use first the Theorem 3 of \cite{CV} to prove that
$f''(x) = ax^{\gamma}$. Then, again we compare the decomposition of Lemma \ref{decomplemma2} and the decomposition of Theorem  \ref{tgr} to get our formulas.
$\Box$
\begin{corollary}\label{coro1} Let $u$ be common utility function and let $f$ be its convex conjugate, $f''(x)=ax^{\gamma}$, where $a>0$ and $\gamma\in\mathbb{R}$. Then for $x>\underline{x}$ there exists an $u$-asymptotically optimal strategy if and only if the conditions of Theorem \ref{exist} are verified for both processes $L$ and $\tilde{L}$.
Furthermore, if $c\neq 0$ then 
$$\phi^*_t=-A_{t}(\tau)\,\frac{ \beta^*_t(\tau) \,( Z_{t-}^*( \tau ))^{\gamma+1}}{S_{t-}\, }$$
with 
$$A_{t}(\tau )= \,\alpha_{\gamma}(x) \,\frac{\mathbb{E} (\,[z^*_{T}( \tau )/z^*_{t}( \tau ) ]^{\gamma +2}\,| \,\tau \,)}{ \mathbb{E}(\,[Z^*_{T}( \tau ) ]^{\gamma +2}\,)}$$
If $c=0$, then we have the same formula for $\phi^*_t$ with replacement of $\beta^*_t(\tau)$ by $\alpha^* _t(\tau)$ given in Theorem \ref{optst1}.
In addition, $\phi^*$ is optimal as soon as $\gamma\neq -1$. 
\end{corollary}

\vskip 0.2cm 
\it Proof \rm According to the Theorem \ref{exist}, under the assumptions (\ref{cdsec1}), (\ref{cdsec2}) and (\ref{cdsec3}), the Levy model associated with $L$ has an $f$-divergence minimal equivalent martingale measure which preserves the Levy property and is scale invariant. The same is true for the Levy model associated with $\tilde{L}$.
Then, the existence of $f$-divergence EMM's for change-point model follows from Theorem \ref{thc} and the formulas for strategies follow directly from Theorem \ref{optst1}.$\Box$\\

\begin{rem} From Corollary \ref{coro1} and Theorem \ref{strat} we can see the following. Let $\psi ^*$ and $\tilde{\psi}^*$ be u-optimal strategies for the exponential Levy models $L$ and $\tilde{L}$ respectively. Then the u-optimal strategy for corresponding change-point model can be written as
$$\phi ^*_t= B_t(\tau)\psi ^*_t \ind _{\{ \tau >t\}} + \tilde{B}_t(\tau)\tilde{ \psi}^*_t  \ind _{\{ \tau \leq t\}}$$
where
$$B_t(\tau)= (c^*(\tau))^{\gamma+1} \frac{\mathbb{E} (\,[\zeta ^*_{\tau} ]^{\gamma +2}\,| \,\tau )\,\mathbb{E} (\,[\tilde{\zeta}^*_{T-\tau} ]^{\gamma +2}\,| \,\tau \,)}{ \mathbb{E}(\,[Z^*_{T}( \tau ) ]^{\gamma +2}\,)}$$
$$\tilde{B}_t(\tau)= (c^*(\tau))^{\gamma+1}(\frac{\zeta_{\tau}}{\tilde{\zeta}_{\tau}})^{\gamma+1} \frac{\mathbb{E} (\,[\tilde{\zeta}^*_{T} ]^{\gamma +2}\, \,)}{ \mathbb{E}(\,[Z^*_{T}( \tau ) ]^{\gamma +2}\,)}$$
When u is exponential utility, and, hence, $\gamma = -1$, we see that $B_t(\tau)=
\tilde{B_t(\tau)}=1$ and  the optimal strategy $\phi^*$ can be obtained by pasting together two optimal strategies  $\psi ^*$ and $\tilde{\psi}^*$ at $\tau$. In addition, in this case $\phi ^*$ is already adapted with respect to progressively enlarged filtration.
\end{rem}
\noindent  \bf{Example}:\,\,\it Optimal strategy for Black-Scholes model with change point and exponential utility. \rm
 As before, we now want to apply the results when $L$ and $\tilde{L}$ define Black-Scholes type models. Therefore, we assume that $L$ and $\tilde{L}$ are continuous Levy processes with characteristics $(b,c,0)$ and $(\tilde{b},c,0)$ respectively. Let $\tau$ be a random variable  bounded by $T$ which is independent from $L$ and $\tilde{L}$ .
Then the asymptotically optimal strategy from the point of view of maximization of exponential utility $u(x) = 1 - \exp (-x)$ will be :
$$\phi ^*_t = - \frac{\beta _t}{S_{t-}}=   \frac{(b+ c/2) \ind_{[\![0,\tau]\!]}(t)+(\tilde{b}+c/2)\,\ind_{]\!]\tau,+\infty[\![}(t)}{c S_{t-}}$$

%%%%%%%%%%%%%%%%%%%%%%%%%%%%%%%%%%%%%%%%%%%%%%%%%%%%%%%%%%%%%%%%%%%%%%%%%%%%%%%
\noindent \section{Acknowledgements} \rm  
The authors would like to thank  Referees of this article for useful remarks and comments.\\\\
This work is supported in part by ECOS project M07M01  and by ANR-09-BLAN-0084-01 of the Department of Mathematics of Angers's University. 

%%%%%%%%%%%%%%%%%%%%%%%%%%%%%%%%%%%%%%%%%%%%%%%%%%%%%%%%%%%%%%%%%%%%%%%
%%%%%%%%%%%%%%%%%%%%%%%%%%%%%%%%%%%%%%%%%%%%%%%%%%%%%%%%%%%%%%%%%%%%%%%

\end{document}